\documentclass{ar-1col}

\usepackage[comma]{natbib}
\usepackage{amssymb,amsmath}
\usepackage{url,xcolor}
\jname{Ann. Rev. Fluid Mech.}
\jvol{AA}
\jyear{YYYY}
\doi{10.1146/((please add article doi))}

\begin{document}
\newcommand{\sm}[1]{{\color{red} #1}}
\newcommand{\fig}[1]{{\color{blue} Figure: #1}}
\newcommand{\Pe}{\mbox{Pe}}
\newcommand{\eb}{\mathbf{e}}
\newcommand{\Ub}{\mathbf{U}}
\newcommand{\nb}{\mathbf{n}}
\newcommand{\dd}{\mathrm{d}}
\newcommand{\totd}[2]{\frac{\mathrm{d} #1}{\mathrm{d} #2}}
\newcommand{\pard}[2]{\frac{\partial #1}{\partial #2}}
\newcommand{\rb}{\mathbf{r}}
\newcommand{\ci}{\mathrm{i}}
\newcommand{\ee}{\mathrm{e}}
\newcommand{\ub}{\mathbf{u}}
\newcommand{\vb}{\mathbf{v}}
\newcommand{\Vb}{\mathbf{V}}
\newcommand{\sigmab}{\boldsymbol{\sigma}}
\newcommand{\Omegab}{\boldsymbol{\Omega}}
\newcommand{\noteTOAR}[1]{{\color{blue}{}}}
\newcommand{\change}[1]{{\color{black} #1}}

\markboth{S. Michelin}{Self-propulsion of chemically-active droplets}

\title{Self-propulsion of chemically-active droplets}

\author{Sébastien Michelin
\affil{LadHyX, CNRS, Ecole Polytechnique, Institut Polytechnique de Paris, 91128 Palaiseau Cedex, France; email: sebastien.michelin@ladhyx.polytechnique.fr}}

\begin{abstract}
Microscopic active droplets are able to swim autonomously in viscous flows: this puzzling feature stems from solute exchanges with the surrounding fluid via surface reactions or their spontaneous solubilisation, and the interfacial flows resulting from these solutes' gradients. Contrary to asymmetric active colloids, these isotropic droplets swim spontaneously by exploiting the nonlinear coupling of solute transport with self-generated Marangoni flows, which is also responsible for secondary transitions to more complex individual and collective dynamics. Thanks to their simple design and their sensitivity to physico-chemical signals, they are fascinating physicists, chemists, biologists and fluid dynamicists alike to analyse viscous self-propulsion and collective dynamics in active matter systems, to develop synthetic cellular models  or to perform targeted biomedical or engineering applications. I review here the most  recent and significant developments of this rapidly-growing field, focusing on the mathematical and physical modelling of these intringuing droplets, together with its experimental design and characterisation.

\end{abstract}

\maketitle

\section{Introduction}\label{sec:intro}
Swimming at microscopic scales is dominated by the effect of viscosity, thus presenting  many puzzling characteristics to our own inertial experience of swimming~\citep{Childress1981,Lauga2009}. Overcoming the conceptual and technical challenges associated with the observation, modeling and manufacturing of micro-swimmers has mobilized biologists, physicists and applied mathematicians since the early observation and modeling of swimming bacteria~\citep{Berg1973,Lighthill1976}. Over the last 15 years, designing synthetic self-propelled systems has become a central engineering challenge, e.g. to perform  biomedical or environmental  tasks~\citep{Nelson2010,JuradoSanchez2018}. It is also driven by a marked fundamental interest for the collective dynamics, self-organisation and control of microscopic systems,  an ever growing field at the cross-road between physics, fluid dynamics and biology, commonly referred to as ``active matter''~\citep{Marchetti2013}. In contrast with  macroscopic  flocks of birds or schools of fish, microswimmer suspensions offer interesting examples of passive interactions, featuring no agent-based decision-making or complex cognitive systems. Hydrodynamics plays here a critical role, as the suspending fluid mediates one of the main interaction routes between different individuals, across dimensions much larger than their body size.

Drawing inspiration from biological systems, the first synthetic micro-swimmers strictly followed bio-inspired designs, such as beating artificial flagella and cilia~\citep{Dreyfus2005,Vilfan2009}. But miniaturisation, assembly and powering at such small scales entail many difficulties, calling for simpler and novel designs. Spontaneous self-organisation also requires  independent individual swimmers, and as such precludes directional macroscopic forcing used in earlier proof-of-concept systems~\citep{Zhang2009}.

 Instead of using miniaturised energy storage and  actuation of deformable structures, chemically-active swimmers convert chemical energy stored internally or in their environment into interfacial flows along their surface, thus prescribing mechanical stresses on the surrounding fluid to achieve self-propulsion~\citep{Moran2017}. Their development is currently driven by their various potential applications, including targeted cargo  or drug delivery~\citep{Kagan2010}, chemical analysis and sensing~\citep{Yi2016} or decontamination~\citep{JuradoSanchez2014}. The most popular designs  consist of coated micro-particles, that catalyse surface chemical reactions (e.g. H$_2$O$_2$-decomposition, enzymatic reactions), and draw their mobility from \emph{phoretic} flows generated from the resulting local chemical gradients~\citep{Paxton2004,Duan2015}.  Yet, those  present significant drawbacks. Being typically micron-size or smaller, their motion is fundamentally Brownian. Although they do not present any moving or miniaturised components, their controlled and reproducible manufacturing is based upon complex physico-chemical techniques (e.g. deposition) and their propulsion requires a built-in asymmetry. 
 
\begin{figure}
\begin{center}
\includegraphics[width=\textwidth]{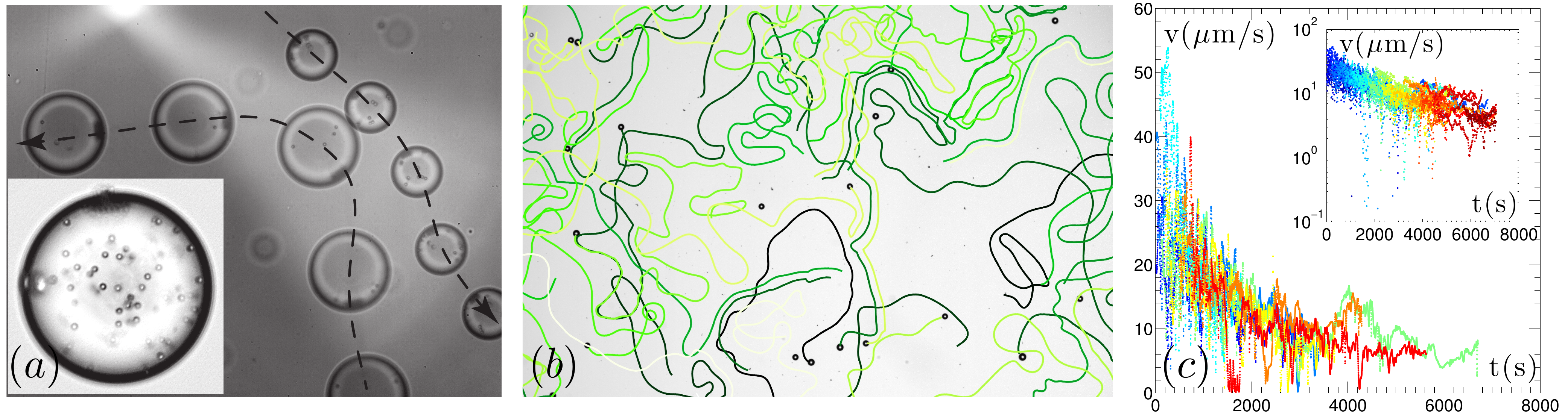}
\caption{Self-propulsion of chemically-active droplets. (a) Individual and (b) collective trajectories of swimming water droplets in monoloein-saturated squalane oil, and evolution in time of the droplet velocities. 
Panels (a) and (c) adapted with permission from~\citet{Izri2014}, copyright (2014) American Physical Society; photo in panel (b) provided by O. Dauchot, M. Van der Linden \& G. Durey, Gulliver lab, CNRS-ESPCI Paris.}
\label{fig:intro}
\end{center}
\end{figure}

Chemically-active droplets are based on the same paradigm, i.e. the combination of a physico-chemical activity with the local forcing of interfacial flows by chemical gradients, but overcome several of these issues. As simple as pure water droplets slowly-dissolving in a surfactant-saturated oil phase, they  swim for hours (Figure~\ref{fig:intro}) and have attracted a rapidly-growing interest since the first experimental reports of their motion~\citep{Izri2014,Maass2016}. Their fluid nature  combined with their ability to encapsulate complex internal chemistry allow them to reproduce several cell-like behaviors, e.g. migration, fusion or division~\citep{Toyota2006,Browne2010,Banno2013}. They can also be used for cargo delivery~\citep{Li2018} or metal extraction~\citep{Ban2014}. Despite their fascinating simplicity, understanding the fundamental principles of their self-propulsion and interactions has nevertheless proved significantly more challenging. Isotropic by design, they must break spontaneously a spatial symmetry in order to self-propel, exploiting a non-linear coupling of their physico-chemical dynamics to the resulting internal and external fluid motions~\citep{Morozov2019}. This intrinsic non-linearity and a dual sensitivity to their hydro-chemical environment enables complex, non-Brownian, individual and collective motion of active droplets (Figure~\ref{fig:intro}), but is also responsible for a significant gap between their modeling and that of their rigid phoretic counterparts, for which chemical transport is mostly diffusive and decoupled from hydrodynamics.

In contrast with other Marangoni-driven self-propelled systems~\citep[e.g. camphor boats][]{Tomlinson1862}, chemically-active swimming droplets appeared in microfluidics labs only very recently, with most  realisations reported over the last ten years. Motivated by such rapid experimental developments and by the complexity and variety of the physico-chemical processes involved, their theoretical understanding is now following suit, with a wealth of analysis, simulations and modeling proposed over the last few years only. The present review proposes a critical panorama of this dynamic  research field, with a synthetic presentation of our current fundamental understanding of these fascinating yet complex systems, interlinked with the evolving state-of-the art of their design and observations. A specific focus will be placed on understanding their distinctive features with respect to other chemically-active swimmers, namely the nonlinear hydro-chemical couplings at play and their consequences. Finally, we also provide an overview of the main ongoing challenges, from a fluid dynamics viewpoint. 

As a result, this review is organised as follows. After a brief reminder in Sec.~\ref{sec:microswimmer} of the specific constraints and minimal ingredients for  microscopic self-propulsion in a fluid medium, Section~\ref{sec:lab} presents a synthetic overview of the main experimental designs and characteristics. The fundamental hydro-chemical mechanism at the heart of their spontaneous self-propulsion and its modelling are presented in Sec.~\ref{sec:modeling_ind}, together with the emergence of complex individual motions. Section~\ref{sec:interactions} is a review of the different  interactions of active droplets, with their environment and amongst each other, a key step toward modelling the large scale collective dynamics of this new type of active matter. Finally, the present review concludes on some of the prominent ongoing and future challenges that must be overcome in our quest to understand the dynamics of these intriguing droplets (Sec.~\ref{sec:perspectives}).

\section{Self-propulsion principles of chemically-active droplets}\label{sec:microswimmer}
\subsection{Swimming at the micron scale}
Chemically-active droplets in the lab have typical radius $R\sim 10-100\mu$m and velocity ${V\sim1-100\mu\textrm{m.s}^{-1}}$. For such {microswimmers}, the \emph{Reynolds number} is small and inertial effects are negligible: fluid flows within and around the droplets are therefore described by the Stokes equations~\citep{Kim1991}. The last 50 years have led to much progress in understanding and modelling viscous propulsion of various swimming micro-organisms~\citep[see the extensive and excellent reviews by][]{Lauga2009,Pak2015}, and many of its features are applicable to any microswimmers, including swimming droplets and other active colloids.  We here only outline some of its most distinctive features:

\begin{marginnote}
\entry{Reynolds number}{$\mbox{Re}=\rho V R/\eta$ (with $\eta$ and $\rho$ the fluid viscosity and density) measures the relative effect of inertia and viscosity on the fluid motion.}
\end{marginnote}
\begin{enumerate}
\item{The swimming speed is a \emph{linear} function of the \emph{instantaneous} forcing exerted by the  microswimmer on the surrounding fluid: there is no history effect and swimming ceases as soon as the  forcing disappears.}
\item{Net motion requires a \emph{spatially-asymmetric} and \emph{temporally-irreversible} forcing.}
\item{Inertia is negligible and a strict force/torque balance applies on the microswimmer at all times. In the absence of any external force (e.g. buoyancy), the total hydrodynamic force and torque thus vanish.}
\item{The flow perturbations introduced by a microswimmer decay slowly (algebraically) resulting in long-range interactions with other swimmers and boundaries.  Force-free microswimmers are  characterised by $O(1/r^2)$ hydrodynamic signature (stresslet).}
\end{enumerate}

\subsection{Generating interfacial stresses and flows}
Viscous propulsion critically relies on the ability to force the surrounding fluid into motion irreversibly. Swimming cells and other biological swimmers achieve this by performing successions of shape changes (known as \emph{strokes}), most commonly through the beating of active filaments (flagella, cilia) anchored on the cellular membrane~\citep{Brennen1977,Lauga2016arfm}. This is also a central strategy of many artificial microswimmers designed in the lab~\citep[e.g.][]{Nelson2010,Palagi2018}. Chemically-active particles or droplets exploit instead physico-chemical gradients and short-range interactions with the surrounding fluid to generate interfacial flows along their boundaries~\citep{Anderson1989}.

The surface of a swimming droplet separates two immiscible fluids containing dissolved ionic or neutral solutes and is most often ladden with large \emph{surfactants}. 
\begin{marginnote}
\entry{Surfactants}{Large molecules featuring a hydrophilic head and hydrophobic tail; they adsorb preferentially at oil-water interfaces, lowering their interfacial stress (\emph{surface tension}).}
\end{marginnote}
The interaction of the different molecular species (solute, solvents and surfactants) is incredibly complex. Yet, because such interactions are limited to the immediate vicinity of the interface, whose thickness is typically orders of magnitude smaller than the droplet's size, these molecular interactions can be accounted for in the mesoscopic point of view of  fluid mechanics  through (i) an \emph{interfacial stress} $\boldsymbol\tau_s=\nabla_s\gamma$, with $\gamma(\rb)$ the {surface tension}, and (ii) a \emph{phoretic slip} velocity $\ub_s$~\citep{Anderson1989}, so that boundary conditions at the surface of the droplet for  the  velocity and stress of the outer and inner fluids, $\ub^{(o,i)}$ and $\boldsymbol\sigma^{(o,i)}$, write
\begin{equation}
\left[\sigmab^{(o)}-\sigmab^{(i)}\right]\cdot\nb=-\boldsymbol\tau_s,\qquad \ub^{(o)}-\ub^{(i)}=\ub_s.
\end{equation} 
 Both $\ub_s$ and $\boldsymbol\tau_s$ are directly linked to local physico-chemical gradients at the droplet interface and drive  net \emph{phoretic} and/or \emph{Marangoni} flows around the droplet, providing the mechanical forcing (\emph{mobility}) that propels the droplet through the viscous fluid.
\begin{marginnote}
\entry{Phoretic flows}{Slip-driven interfacial flows resulting from surface physico-chemical gradients.}
\entry{Marangoni flows}{Stress-driven interfacial flows resulting from surface tension gradients at fluid-fluid interfaces.}
\end{marginnote}

\subsection{Minimum ingredients for phoretic/Marangoni propulsion}
Phoretic and Marangoni migration of passive colloids resulting from externally-imposed physico-chemical gradients is a classical problem~\citep{Anderson1989}. The distinguishing feature of self-propelled active colloids lies in their ability to create and act on such local gradients themselves through a surface \emph{activity} of various origins (see Section~\ref{sec:lab}).

\begin{figure}[h]
\begin{center}
\includegraphics[width=.9\textwidth]{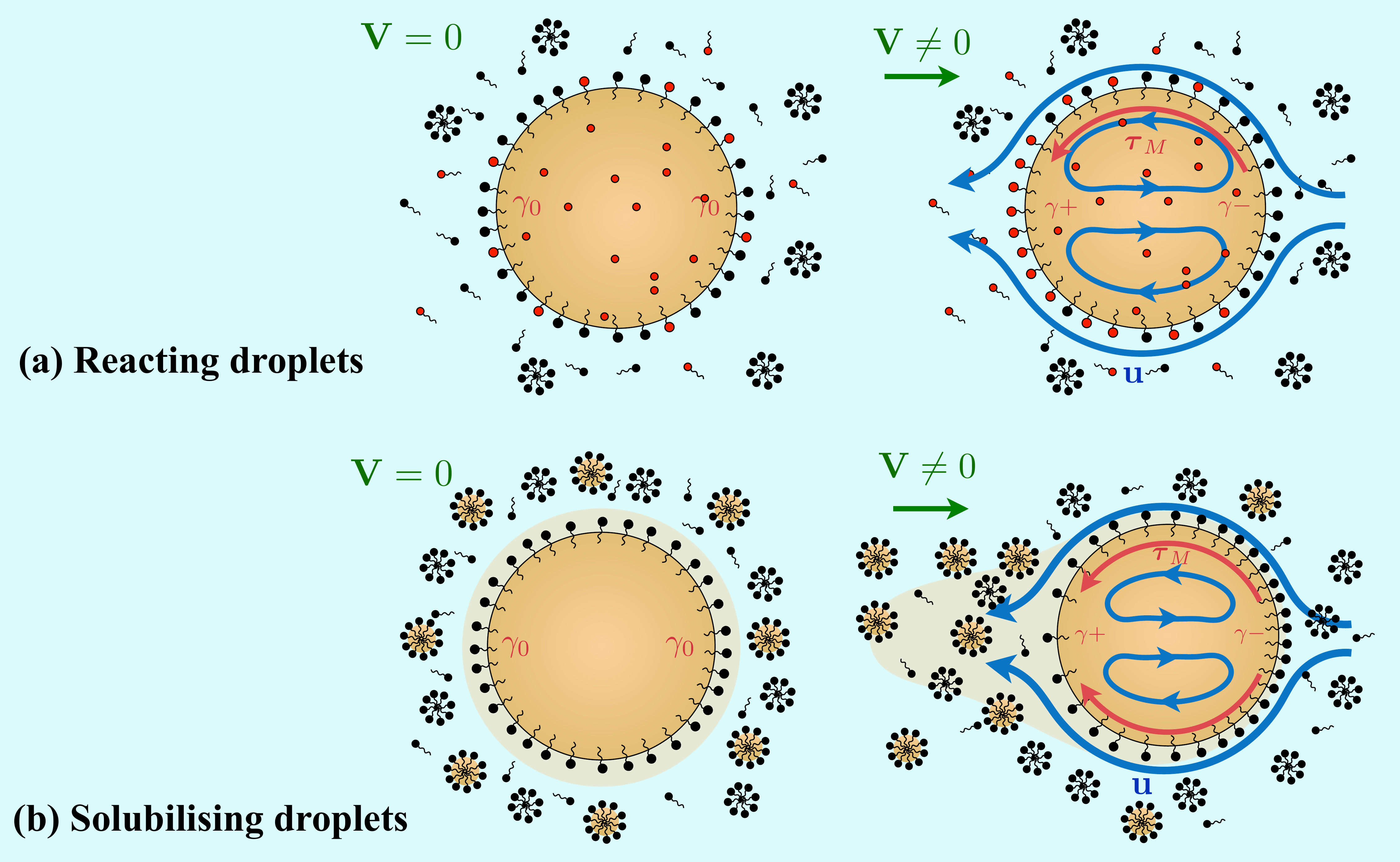}
\caption{ (Left) Isotropic non-motile equilibrium and (right) symmetry-breaking and self-propulsion of two canonical examples of chemically-active droplets. (a) \emph{Reactive droplets:} a chemical (red) produced inside the droplet reduces the tensio-active properties of the adsorbed surfactant molecules. Regions of converging surface flows are rich in altered (red) surfactant molecules (increased surface tension) while diverging surface flows replenish the surface in pristine (black) surfactant (reduced surface tension), thus enhancing the fluid and droplet motions. (b) \emph{Self-solubilising droplets:} the droplet's fluid is transferred near the surface into swollen micelles, reducing the local surfactant concentration. The accumulation of swollen micelles in the wake of the moving droplet thus maintains the interfacial stress (red arrows) and Marangoni flow (blue arrows) driving the droplet's motion. \noteTOAR{this figure was created by me for this article.}}
\label{fig:mechanism}
\end{center}
\end{figure}

In addition to a dual activity and mobility, self-propulsion requires breaking spatial symmetry. Being intrinsically isotropic, chemically-active droplets can not rely on built-in asymmetries, a stark contrast with rigid phoretic particles~\citep{Moran2017}. Active droplets exploit instead the convective transport of the product of their activity to maintain/reinforce interfacial propelling flows in a spontaneous symmetry-breaking (Figure~\ref{fig:mechanism}).

\section{Chemically-active droplets in the lab}\label{sec:lab}
This review is dedicated to colloidal droplets that self-propel due to self-generated interfacial flows resulting from surface physico-chemical gradients (e.g. solute, temperature...). Many recent exemples of such chemically-active droplets have been proposed in the lab, and  can be broadly classified into two main categories, namely (i) \emph{reacting droplets} and (ii) \emph{solubilising droplets}~\citep{Herminghaus2014,Maass2016}. 
The former typically exploit chemical reactions to alter the structure of surfactant molecules at their interface, modifying their tensio-active properties.  In contrast, the latter rely on micellar dissolution into their surfactant-saturated environment.  Both mechanisms affect the droplet's surface tension, and may therefore lead to self-induced Marangoni flows and propulsion of the droplet.
 
\subsection{Reacting droplets}
Such droplets modify their surface tension distribution using internal or surface reactions to produce or alter surfactant molecules covering their surface. The dynamics of internal reactions are essentially independent from the outer environment, in contrast with surface reactions where chemical reagents are transported within each phase and brought into contact at the interface only. In that case, propulsion itself is controlled by the surfactant concentration at and near the droplet surface. 
The possibility to encapsulate complex chemical reactions makes them particularly well-suited for targeted applications. Popular examples include:
\begin{itemize}
\item{\emph{Surfactant production}, e.g. of a cationic surfactant inside an oil droplet via a catalytic cyclo-addition~\citep{Kasuo2019};}
\item{\emph{Surfactant bromination}: the tensio-active properties of the surfactant (e.g. monoolein) are altered by the bromination of an unsaturated bond in its hydrocarbon chain. Bromine can be loaded initially inside the droplet~\citep{Suematsu2019} or produced internally via steady or oscillatory Belouzhov-Zhabotinsky (BZ) reactions, depending on the droplet composition~\citep{Thutupalli2013,Thutupalli2011,Suematsu2016}.}

\item{\emph{pH-controlled hydrolisation} of an ester-containing surfactant in contact with the aqueous phase  to produce a more active surfactant complex~\citep{Banno2011,Ban2013}. The reaction kinetics, and thus the droplet's propulsion properties (e.g. duration, speed) may be tuned via the pH of the aqueous outer phase. 
  }
\end{itemize}

\subsection{Solubilising droplets (spontaneous emulsions)}
In this second popular route to self-propulsion, active droplets dissolve spontaneously into the surfactant-saturated outer phase, in a mechanism known as \emph{micellar dissolution}~\citep{Herminghaus2014,Maass2016}. This somewhat simpler physical chemistry has been particularly attractive for fundamental studies of droplet motion and interactions. The mechanism is illustrated below for oil-in-water droplets, but a similar mechanism is also available to water-in-oil droplets.

\begin{marginnote}
\entry{Critical Micellar Concentration (CMC)}{Maximum  concentration of surfactant monomers in a given solvent; additional surfactants form supra-molecular aggregates (micelles) to ``shield'' their head or tail  from the solvent.}
\end{marginnote}

Above the \emph{Critical Micellar Concentration}, some of the droplet's oil is spontaneously transferred into empty micelles present in the vicinity of the droplet's surface, together with additional surfactant monomers, in order to form larger \emph{swollen} micelles. This process therefore alters the local surface tension, and constitutes the gradient-inducing \emph{activity} required for propulsion. Micelle swelling and droplet dissolution are complex physico-chemical processes beyond the scope of this review, but follow either (i) a micellar pathway where micelles are filled directly at the droplet's surface, or (ii) a molecular pathway where swollen micelles are formed from dissolved oil molecules present in the surface vicinity~\citep[see][for more details]{Herminghaus2014}. 

A popular system consists in the dissolution of 4-pentyl-4'-cyanobiphenyl (5CB) oil droplets in a saturated aqueous solution of tetradecyltrimethylammonium bromide (TTAB) surfactant~\citep{Peddireddy2012,Seemann2016}. This liquid crystalline oil is nematic at room temperature. Yet, nematic ordering is not necessary to the droplet propulsion: swimming is maintained above the nematic-to-isotropic transition~\citep[][]{Kruger2016b}, and is also observed for non-nematic LC oils~\citep[such as CB15][]{Hokmabad2021} or non-LC oils such as diethyl phtalate (DEP)~\citep{Moerman2017,Izzet2020} or bromodecane~\citep{Cheon2021}. Nematic ordering further provides an interesting visualisation tool of  the droplet's internal structure during the motion~\citep{Herminghaus2014,Thutupalli2018}.

A similar micellar dissolution also powers the propulsion of water droplets in surfactant-saturated oil~\citep[e.g. monoloein-saturated squalane\footnote{This is also the combination used for BZ-droplets~\citep{Thutupalli2011}: micellar solubilisation may therefore coexist with chemically-driven propulsion for this system~\citep[][]{Suematsu2021}.}, see][]{Izri2014,deBlois2019,Suda2021}. 
More complex physico-chemical routes  have also been explored for the droplet's spontaneous dissolution such as the shedding of larger fluid vesicles at the back of the propelling droplet~\citep{Toyota2006,Toyota2009}.

\subsection{Symmetry-breaking and other general Marangoni swimmers}
A key feature of all these examples is the intrinsic isotropy of active droplets, that must rely on the nonlinear solute transport dynamics to break symmetry. Other swimmers rely on the same physical principles, e.g. isotropic Marangoni surfers propelling at a fluid interface~\citep{Ender2021a,Boniface2019}. Yet, such systems also show fundamental differences: Marangoni flows are generated here along the entire fluid interface instead of the sole swimmer's surface, playing a major role in its propulsion~\citep{Boniface2021}.

\section{Self-propulsion of an isolated droplet: fundamental mechanisms and modeling}
\label{sec:modeling_ind}

\subsection{Modeling  chemically-active droplets}
Given such experimental diversity, it should be no surprise that modeling the self-propulsion of chemically-active droplets remains an important challenge and follows different approaches. We summarise below their common characteristic features, together with a simple ``reference'' model and possible variants.
\subsubsection{Hydro-chemical transport}
Self-propulsion of chemically-active droplets emerges from the nonlinear coupling of (i) the transport of the different chemical solutes around the droplet and (ii) the viscous fluid motion resulting from the interfacial forcing introduced by these solutes' gradients. Originating in the slow diffusion of the large solutes, this nonlinear hydro-chemical coupling is  essential to the emergence of directed self-propulsion of intrinsically-isotropic droplets.

The detailed propulsion mechanism is still not completely understood in a large part due to the complexity of the  physico-chemical processes, and its variety across experimental systems. Yet, key common features are found in most cases, including a dual activity and mobility. The modelling of the active droplet and of its self-propulsion can be broadly rationalised along two main classes of mechanisms and models:
\begin{enumerate}
\item{\textbf{\emph{Interfacial} hydro-chemical dynamics}: This mechanism is best-suited for the description of reacting droplets, for which a reaction takes place \emph{at} the interface~\citep{Thutupalli2011,Schmitt2013}: the \emph{activity} of the droplet may then consist in the alteration/degradation of pristine surfactant with a consumption rate directly related to its local concentration at the interface. The different surfactant molecules are transported along the interface by the solute-induced Marangoni flows.  Locally-converging surface flows drive an outward radial flow into the bulk, reducing the influx of pristine surfactant from the saturated solvent and its local surface concentration. By raising the local surface tension, this enhances the initial flow perturbation in a self-reinforcing mechanism at the origin of the spontaneous onset of propulsion (Figure~\ref{fig:mechanism}a). By effectively introducing a spatially-dependent activity directly driven by the surface flow divergence, this tightly-coupled hydro-chemical dynamics resembles the similar problem of surface-mobile enzymes~\citep{DeCorato2020b}.

A simple hydro-chemical model of such mechanism writes for the reaction-advection-diffusion dynamics of the relative \emph{surface} concentration of pristine surfactant~\citep{Thutupalli2011,Schmitt2013} with the  convecting surface flow  obtained by solving Stokes' equations inside and outside the droplet in linear response to the Marangoni forcing~\citep{Morozov2019b,Schmitt2013,Schmitt2016}.}

\item{\textbf{\emph{Bulk} hydro-chemical dynamics}: In contrast, the {activity} of self-solubilizing droplets consists in the exchange of chemical species between the droplet's surface and the outer fluid.  
Resulting surface tension inhomogeneities drive  Marangoni flows (\emph{mobility}) and a bulk transport of these chemical species. Such micellar solubilisation can be broadly interpreted equivalently as the consumption of a surface-tension-reducing solute (empty micelles and adsorbed monomers) or the production of a surface-tension-increasing one (swollen micelles). It can also be seen as an effective reduction of the local CMC (and surfactant monomer concentration) due to the presence of small amounts of dissolved oil molecules transported by the flow in the droplet's vicinity~\citep{Izzet2020}. Focusing on the first interpretation, a small droplet displacement thus results in the convective solute transport and accumulation in the droplet's wake; the Marangoni flow driven by the resulting increased surface tension is then able to propel the droplet forward (Figure~\ref{fig:mechanism}b).}
\end{enumerate}

More complete models also couple the interfacial and bulk transports through the description of the adsorption/desorption dynamics of surfactant molecules from the interface~\citep{Rednikov1994b,Yoshinaga2012,Morozov2020}: the two classes presented above can then be viewed, at least qualitatively, as limit cases of the sorption kinetics of the surfactant species. As temperature changes also alter surface tension, most of the above discussion of chemically-active droplets  remains applicable for heat-releasing droplets~\citep{Rednikov1994d}.\\

All models share common and essential characteristics  for spontaneous  propulsion:

\begin{summary}[Propulsion mechanism of chemically-active droplets]
\begin{enumerate}
\item{\emph{Destabilisation}: Local physico-chemical surface gradients (e.g. temperature or solute content) generate an interfacial flow, and the associated convective chemical transport enhances the initial inhomogeneity;}
\item{\emph{Threshold}: Self-sustained convective transport of the field must dominate over chemically- or diffusion-driven relaxation toward equilibrium.}
\end{enumerate}
\end{summary}

\subsubsection{Marangoni vs. phoretic forcing}
For simplicity, the presentation above focused specifically on Marangoni-driven flows. Yet,  linear modelling of the surface tension dependence is only valid for low surfactant coverage. Furthermore, phoretic slip may also develop at a fluid-fluid interface in the presence of a solute concentration, and the balance between the two processes is determined by the detailed molecular interactions, the surfactant coverage density or the viscosity ratio. A more general approach thus appears necessary and consists in retaining both drivings for the interfacial flow~\citep{Morozov2019b}. In that regard, results on phoretic propulsion at finite or large $\Pe$ (obtained from a somewhat simpler model that does not require describing the internal fluid's motion) are also likely relevant to the droplet problem~\citep{Yariv2016,Yariv2017b,Saha2021,Chen2021}. The resulting dynamics in either case are indeed essentially the same for moderate $\Pe$~\citep{Michelin2013c,Izri2014}, but significant differences have been identified for larger $\Pe$ \citep[e.g. subsequent bifurcations and transition to chaos, see][]{Morozov2019b,Lin2020}.

\subsubsection{Relaxation dynamics}
Individual models differ in their description of the solute's relaxation. In interfacial mechanisms, this relaxation originates in surface diffusion, degradation or desorption from the surface~\citep{Thutupalli2011,Schmitt2013,DeCorato2020b}. For bulk mechanisms, relaxation stems from a bulk degradation~\citep{Yoshinaga2012} or diffusive / convective transport away from the particle~\citep{Izri2014,Morozov2019,Picella2022}.  The presence (or not) of a bulk relaxation significantly impacts the localisation of the chemical footprint of the droplet, by introducing an effective exponential screening length for the solute concentration, instead of its diffusion-induced algebraic decay away from the droplet. The resulting chemically-mediated interactions of active droplets with surrounding boundaries or other droplets may be significantly altered~\citep{Lippera2020,Lippera2021,Yabunaka2016}.

\subsubsection{Droplet deformability}
Being liquid, active droplets are expected to deform in response to the viscous stresses applied by the inner and outer flows, with relative deformations of the order of $\mbox{Ca}\sim \eta \mathcal{V}/\gamma$ with $\mathcal{V}$, $\eta$ and $\gamma$ the typical relative flow velocity, fluid viscosity and surface tension. The models outlined previously can be directly generalised to account for these, at least in the limit of small deformability~\citep{Rednikov1994c,Yoshinaga2014,Morozov2019}. However, in most experimental situations $\mbox{Ca}\ll 1$ justifies focusing on purely-spherical droplets. 

\subsection{A reference model} 
\label{sec:canonical_model}
Keeping in mind these possible variations, we outline below a simple advection-diffusion model  for the bulk transport of a chemical solute emitted from the surface of the moving active droplet into the outer fluid~\citep{Michelin2013c,Morozov2019b}, that has served as a growingly popular reference for the understanding of these systems. Despite likely shortcomings in the description of the droplet's interfacial physico-chemical properties, its simplicity recently triggered and enabled much fundamental understanding of this rich system~\citep{Saha2021} and investigation of droplet-droplet interactions~\citep{Lippera2020,Lippera2020b}, confinement effects~\citep{Desai2021,Picella2022}, complex unsteady dynamics~\citep{Morozov2019c,Hokmabad2021} or transition to chaotic motions~\citep{Morozov2019b,Hu2019,Chen2021}. 

\subsubsection{Hydrochemical transport equations}
The transport equations write for the inner and outer fluid flows, $\ub^{(o,i)}$ and the concentration perturbation $C$ in the outer fluid, as
\begin{align}
\pard{C}{t}+\ub^{(o)}\cdot\nabla C=D\nabla^2C, \qquad 
\eta_{o,i}\nabla^2\ub^{(o,i)}=\nabla p^{(o,i)},\,\nabla\cdot\ub^{(o,i)}=0.\label{eq:transport}
\end{align}
The droplet has radius $R$, and its activity and Marangoni/phoretic mobility  further impose 
\begin{align}
D\nb\cdot\nabla C=-\mathcal{A},\qquad
\ub^{(o)}-\ub^{(i)}=\mathcal{M}\nb^\perp\cdot\nabla C,&\qquad \nb^\perp\cdot[(\sigmab^{(o)}-\sigmab^{(i)})\cdot\nb+\gamma_C\nabla C]=0 \label{eq:surfbc2}
\end{align}
on the droplet's surface with $\boldsymbol\sigma^{(o,i)}$ the associated Newtonian fluid stress tensor. Here, $\mathcal{A}$ accounts for a fixed-flux emission of solute of diffusivity $D$, $\gamma_C=(\partial \gamma/\partial C)$ is the surface tension sensitivity to solute gradients and $\mathcal{M}$ the phoretic mobility. In unbounded domains, the concentration and flow perturbations  decay far from the droplet (i.e. $C,\,\ub^{(o,i)}\rightarrow 0$ for $|\rb|\gg R$) and the impermeability of the droplet writes $\ub^{(o,i)}\cdot\nb=\Vb\cdot\nb$ on $\mathcal{S}$. The droplet's velocity $\Vb$ is then obtained from the force balance (inertia is negligible at such scales), so that in the absence of any external force
\begin{equation}
\mathbf{F}_h=\int_\mathcal{S}\sigmab^{(o)}\cdot\nb\,\dd S=\mathbf{0}.\label{eq:force}
\end{equation}
Note that these conditions can be modified straightforwardly to account for an external force (e.g. buoyancy) or inhomogeneous chemical and/or hydrodynamic backgrounds.

\subsubsection{P\'eclet number}
 This model features three important non-dimensional parameters:
\begin{equation}
\Pe=\frac{\mathcal{V}R}{D},\qquad \tilde m=\frac{\eta_i\mathcal{M}}{\gamma_CR},\qquad \tilde\eta=\frac{\eta_o}{\eta_i}\cdot\label{eq:ndnumbers}
\end{equation} 
with $\mathcal{V}$ the characteristic magnitude of the activity-driven interfacial flows (e.g. ${\mathcal{V}=\mathcal{AM}/D}$). The P\'eclet number, $\Pe$, is a relative measure of convective-to-diffusive solute transport; the mobility ratio $\tilde{m}$ compares the role of phoretic and Marangoni forcings and $\tilde\eta$ is the viscosity ratio. The limit of a pure droplet is recovered for $\tilde m\ll 1$, and that of a phoretic particle for $\tilde m\gg 1$ and $\tilde\eta\ll 1$~\citep{Morozov2019b}.                                    

The P\'eclet number is also a relative measure of the droplet's activity and/or mobility magnitude, or of the droplet's size. It is sometimes referred to as the Marangoni number, to distinguish it from the same quantity built on the \emph{a priori} unknown droplet velocity $V$ instead of the intrinsic velocity scale $\mathcal{V}$.

\subsection{Symmetry-breaking and the emergence of self-propulsion}
Chemically-active droplets present no intrinsic asymmetry. A non-moving isolated active droplet thus modifies its chemical environment isotropically ($\bar{C}=\mathcal{A}R^2/r$) and its surface concentration remains uniform, resulting in no fluid or droplet motion regardless of $\Pe$. Droplet propulsion thus necessarily involves a pontaneous symmetry-breaking, which stems from the nonlinear hydrochemical coupling introduced by the convective transport.

\subsubsection{Stability of the isotropic state}
\begin{figure}
\begin{center}
\includegraphics[width=\textwidth]{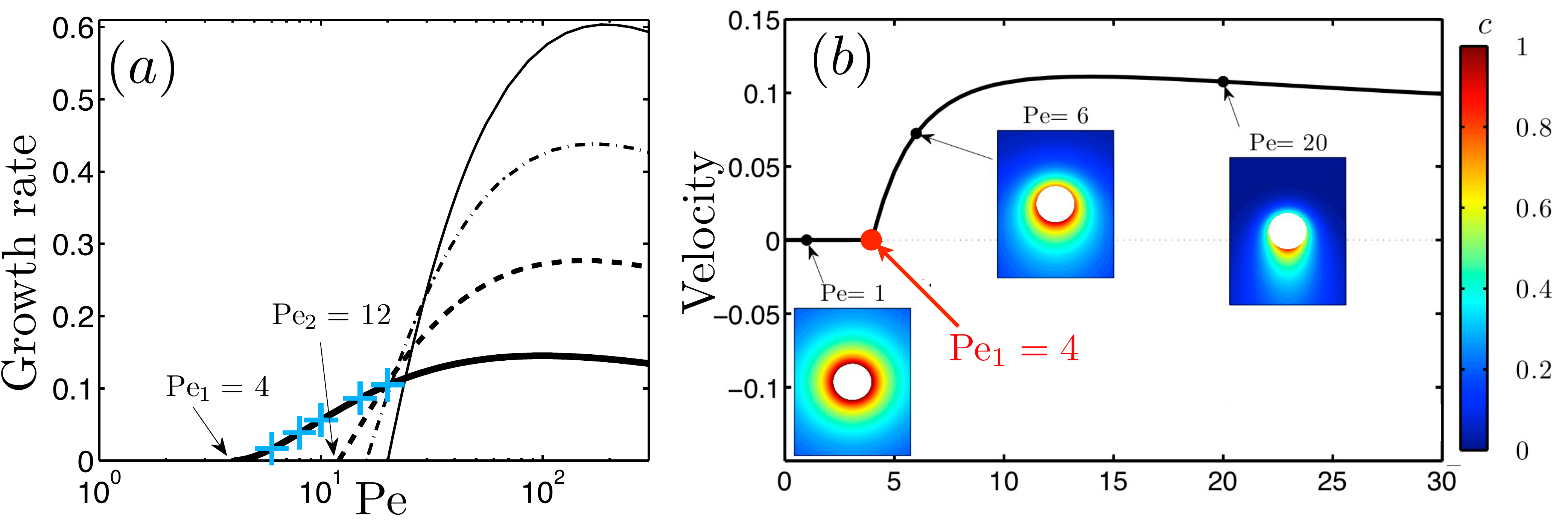}
\caption{Symmetry-breaking and onset of droplet motion. (a) Linear growth rate of the successive unstable modes of the isotropic equilibrium. (b) Evolution with $\Pe$ of the active droplet's velocity and of its chemical footprint, obtained from numerical simulations of the reference nonlinear hydro-chemical model. \change{Self-propulsion occurs for $\Pe\geq \Pe_1=4$.} Panel a adapted  with permission  from~\citet{Michelin2013c};copyright (2013) AIP Publishing LLC. Panel b adapted  with permission  from~\citet{Izri2014}; copyright (2014) American Physical Society.}\label{fig:stab_onset}
\end{center}
\end{figure}

The stability analysis of the reference model (Eqs.~\eqref{eq:transport}--\eqref{eq:force}) around this non-propelling regime confirms the qualitative arguments outlined above. A discrete set of increasing critical thresholds, $\Pe_c^{(n)}$, is identified for force-free droplets, beyond which specific axisymmetric modes of increasing azimuthal order $n$ become successively unstable as diffusion becomes too slow to counteract the amplifying effect of the convective coupling (Figure~\ref{fig:stab_onset}a). The unstable mode associated with the lowest threshold, $\Pe_c^{(1)}$, is dipolar and the only one associated with a net droplet motion; $(\Pe_c^{(n)})_{n\geq 2}$ correspond to flow-stirring modes by a non-moving force-free droplet~\citep{Izri2014,Morozov2019,Morozov2019b,Hokmabad2021}. For tethered droplets, $\Pe_c^{(1)}$ is replaced by an alternative threshold $\Pe_c^*$ corresponding to fluid pumping by a fixed droplet exposed to a net external force~\citep{Rednikov1994b}.

The self-propelling mode always becomes unstable first for spherical droplets: droplet motion toward lower solute concentration and resulting surface flow divergence enhances the destabilising front-back contrast in solute distribution. Significant droplet deformability  promotes the instability of higher-order modes with similar symmetry, potentially modifying the threshold's ordering for larger $\mbox{Ca}$~\citep{Yoshinaga2014,Morozov2019}.

\subsubsection{Near-threshold dynamics and bifurcation characteristics} 
For $\Pe>\Pe_c^{(1)}$, small perturbations of the isotropic state grow exponentially then saturate under the nonlinear effect of hydrochemical coupling (Figure~\ref{fig:stab_onset}b). All directions being \emph{a priori} equivalent, the direction of motion is thus chosen by the initial perturbation~\citep{Saha2021}. Numerical solution of the full non-linear model provides the bifurcation curve indicating the emergence of a new stable branch with $V\neq 0$ above $\Pe_c^{(1)}$ as the isotropic state (with $V=0$) becomes unstable (Figure~\ref{fig:stab_onset}b). Steady self-propulsion is associated with a self-sustained accumulation of solute in the wake of the droplet and the development of a sharp boundary layer at the front of the droplet as $\Pe$ is increased. 

Weakly non-linear analysis shows that $V\sim \Pe-\Pe_c^{(1)}$ at the propulsion onset, instead of the classical square-root behaviour of supercritical pitchfork bifurcations~\citep{Morozov2019}. This in fact points to the singular behaviour of the advection-diffusion problem for small droplet velocity in unbounded domains~\citep{Farutin2021a}: advection is subdominant with respect to diffusion in the near-field, i.e. for $r=O(R)$, but must be retained at leading order in a far-field boundary layer, $r=O(\Pe V)$, resulting in a symmetry-preserving pseudo-quadratic dominant correction of the amplitude equations,  $\displaystyle\totd{V}{t}=\sigma V-K|V|V$~\citep{Velarde1996}. 

The exact nature of the bifurcation thus depends on the details of the model considered: the convective far-field boundary layer is absent when chemical transport is restricted to the droplet surface, or when the chemical field at large distances is screened by bulk degradation~\citep{Yoshinaga2012} or finite domain sizes~\citep{Misbah2021}, thus leading to a classical pitchfork bifurcation. Subcritical transition may also be obtained when other nonlinear corrections are included~\citep[e.g. surface coverage saturation,][]{DeCorato2020b} or when the droplet's activity is concentration-dependent~\citep{Rednikov1994b}.

\subsection{Complex dynamics of isolated droplets}

\begin{figure}
\begin{center}
\includegraphics[width=.99\textwidth]{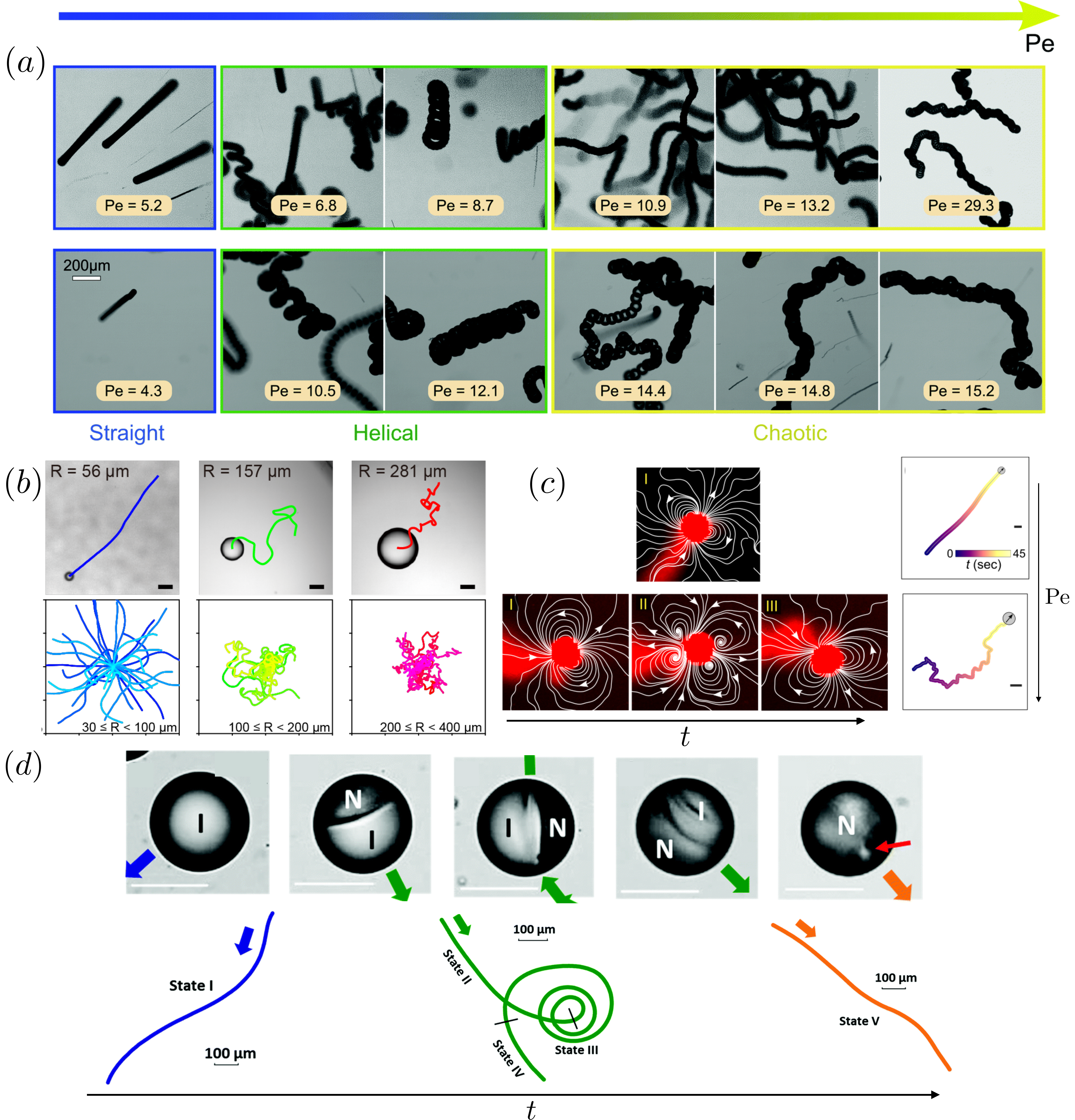}
\caption{\change{(a) Dynamics of CB15 oil droplets in TTAB saturated aqueous solution for increasing surfactant concentration (top) or droplet size (row).} \change{(b) }Swimming dynamics of water droplets of increasing radii in monoolein-saturated solutions of squalane oil. \change{(c)} Experimental visualisation of the solute field (red) and streamlines (white) generated by a self-propelled CB15 oil droplet solubilising in a aqueous solution of TTAB cationic surfactant, as well as its trajectory when $\eta_o$ (and $\Pe$) is modified by adding glycerol to the aqueous phase. \change{(d)} Visualisation of transient dynamical changes of  multicomponent oil droplets triggered by phase separation and internal reordering. \change{Panel a reproduced  from~\citet{Hokmabad2022};CC-BY-3.0.} Panel b reproduced  with permission  from~\citet{Suda2021}; copyright (2021) American Physical Society. Panel c adapted from~\citet{Hokmabad2021};CC-BY-4.0. Panel d adapted   with permission  from~\citet{Wang2021}; copyright (2021) Royal Society of Chemistry. \noteTOAR{permission was purchased on CCC.}}
\label{fig:complextrajectories}
\end{center}
\end{figure}

Balistic motion is observed in specific experimental conditions~\citep[e.g. for $\Pe\gtrsim\Pe_c^{(1)}$,][]{Suga2018,Izzet2020,Hokmabad2021}, but complex motions of chemically-active droplets are commonly observed for more general settings (Figure~\ref{fig:complextrajectories}), e.g. by changing the surfactant's saturation of the outer phase~\citep{Ban2012}, its chemical structure ~\citep{Banno2013}, its affinity to the outer phase~\citep{Hirono2018} or the droplet's internal chemical state~\citep{Suematsu2021}.

\subsubsection{Secondary transitions and mode switching}

Recent experiments provide some insight on these complex dynamics using well-monitored changes in experimental conditions.  Increasing the droplet's size~\citep{Suga2018,Suda2021,Hokmabad2022} or its effective activity~\citep[e.g. solute, surfactant or reactant addition,][]{Dwivedi2021a,Izzet2020,Suematsu2021,Hokmabad2020b}, or reducing the solvent's viscosity~\citep{Hokmabad2021} all effectively increase $\Pe$ and trigger unsteady motion, curved trajectories and erratic changes (Figure~\ref{fig:complextrajectories}\change{a-b}). 

These can be understood as secondary transitions affecting the self-propelled steady state of the reference model (Sec.~\ref{sec:canonical_model}). The dipolar mode is the only unstable one for lower $\Pe$, but higher order modes become successively unstable for increasing $\Pe$, leading to the possible coexistence of several unstable modes with different polar symmetries and associated with significantly different droplet behaviour~\citep{Morozov2019b}. Transitions between motile and non-motile states may be triggered by droplet collisions~\citep[][]{Lippera2020b} or occur spontaneously under the effect of fluctuations or  solute accumulation in the droplet's environment~\citep{Hokmabad2021}: the resulting dynamics is bimodal (``stop-and-go'') as the droplet continuously switches from a (dipolar) propelling mode to a (quadrupolar) stopping/reorienting mode (Figure~\ref{fig:complextrajectories}c).

Droplet shrinking is too slow to impact the propulsion dynamically~\citep{Izri2014},  but enables a quasi-steady scan of a wide range of sizes during a single experiment: large droplets initially display complex random-like behaviour before switching to ballistic  motion and finally stopping altogether as $R$ (and thus $\Pe$) is reduced~\citep{Suga2018}.

\subsubsection{Emergence of chaotic behaviour}
Decorrelation of the droplet's orientational dynamics is observed as $\Pe$ is increased further, with erratic or chaotic-like motion reported experimentally~\citep{Hokmabad2021,Dwivedi2021a} and numerically~\citep{Morozov2019b,Lin2020,Chen2021}. A quantitative insight into possible chaotic transitions has so far been obtained  for constrained geometries: for axisymmetric motion of the droplet, \cite{Morozov2019b} observed the emergence of a symmetric and quadrupolar pumping mode (no propulsion) followed by a chaotic motion for increasing $\Pe$, and showed that such large-$\Pe$ transition is promoted by the presence of a chemically-induced slip at the droplet surface, and is in fact absent for purely Marangoni forcing. Using a two-dimensional ansatz (a ``cylindrical'' droplet), \cite{Hu2019} observed successive and increasingly close transitions leading to a fully chaotic regime.

Yet, these do not fully represent  generic 3D motions:  axially-constrained droplets must fully stop and run into their own chemical wake to change direction~\citep{Morozov2019b}; to be well-posed, 2D analyses require a finite-distance chemical screening, whose size directly controls the motion onset or its characteristics~\citep{Hu2019}, therefore calling for a more complete characterisation of secondary instabilities and chaotic transition.

\subsubsection{Chemically-driven  dynamics}
Complex internal chemistry may also directly trigger intricate individual dynamics, as for the popular BZ droplets for which different behaviours have been observed depending on the exact internal composition and chemical state: (i) slow and persistent motion~\citep[no bromine,][]{Izri2014,Suematsu2021}; (ii) fast and reorienting propulsion~\citep[fixed supply or steady bromine production,][]{Thutupalli2011,Suematsu2019}; (iii) retarded transition between these regimes~\citep{Suematsu2021}; or (iv) large amplitude oscillations~\citep[chemical waves,][]{Kitahata2011,Suematsu2016}.
Multiphase solubilizing droplets provide another example of internal chemical control of the droplet propulsion (Figure~\ref{fig:complextrajectories}d), with phase separation and temporal evolution of the droplet behaviour triggered by solubilisation differences of the droplet components~\citep{Li2018,Li2020,Wang2021}.

\subsubsection{Internal droplet structure (Liquid-crystal droplets)}\label{sec:LC}
Many experiments focus on swimming liquid crystal oil droplets~\citep{Peddireddy2012,Herminghaus2014} or shells~\citep{Hokmabad2019}. While not necessary for self-propulsion, the internal nematic ordering of the droplet can significantly  affect its propulsion~\citep[][]{Kruger2016b,Suga2018}. Non-swimming nematic droplets feature a central defect that is pushed away from equilibrium by the internal flow forcing as swimming develops, thus introducing internal stresses. The emergence of spiral trajectories may then be interpreted as a two-way coupling between an internal polarisation of the nematic ordering and the hydro-chemical dynamics~\citep{Morozov2019c}. More recent experiments have however also demonstrated that internal droplet ordering is not necessary for the emergence of such curved and complex motion, which is also observed for non-nematic droplets~\citep[][]{Hokmabad2022}.

\section{Droplet interactions}
\label{sec:interactions}
\subsection{Chemotaxis and reverse auto-chemotaxis of active droplets}
Active droplets introduce a chemical footprint on their environment due to their exchange of chemical solutes. In turn, chemical inhomogeneities around them induce interfacial stresses; the resulting drift and reorientation provide them with a chemotactic behaviour similar to biological microorganisms~\citep{Berg1975}.

\begin{figure}
\begin{center}
\includegraphics[width=.95\textwidth]{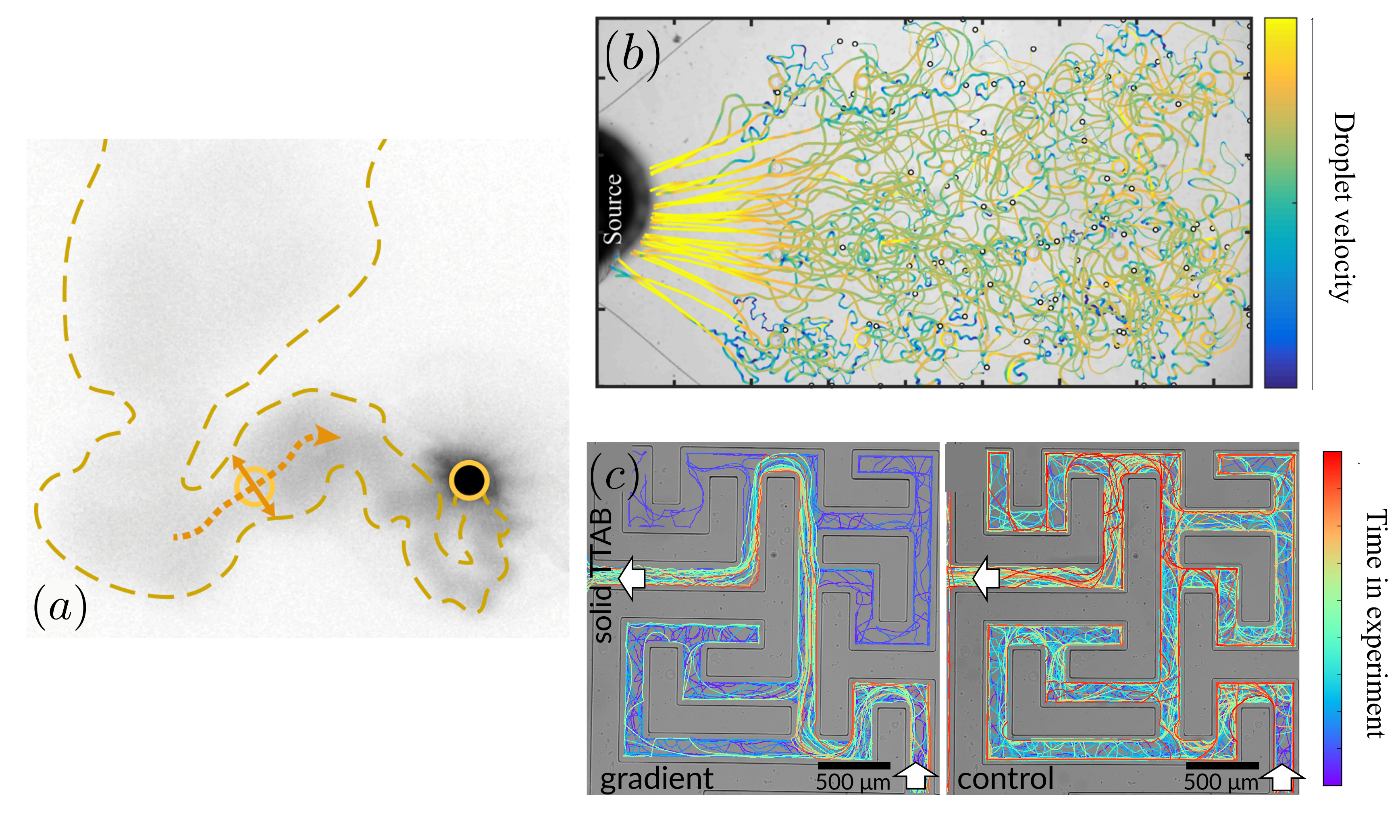}
\caption{(a) Experimental visualisation of the self-avoiding dynamics of a swimming oil droplet as a result of chemo-repulsive interactions with oil-filled micelles. (b) Chemotactic droplets can swim toward a concentrated source of surfactant or (c) find the shortest path to maze's entrance. Panel a adapted from~\citet{Izzet2020}; CC-BY-4.0. Panel b adapted  with permission from~\citet{Jin2018}; copyright (2018) IOP Publishing LTD\noteTOAR{ ; permission was purchased on CCC}. Panel c adapted  with permission  from~\citet{Jin2017}\noteTOAR{ ;PNAS indicated no copyright note was needed}. }\label{fig:chemotaxis}
\end{center}
\end{figure}

The self-propulsion of a solubilising droplet is powered by the capture of some of its internal fluid and additional surfactant monomers from its surface vicinity to form more stable swollen micelles. A local accumulation of swollen micelles or dissolved droplet fluid (e.g. in their wake, \ref{fig:chemotaxis}a) therefore increases the local surface tension and pushes the droplet away from such accumulation regions~\citep[][]{Izzet2020}. As a result, active droplets feature an intrinsic \emph{reverse auto-chemotaxis}~\citep{Maass2016} (i.e. they flee their own wake), and their dynamics is altered by other droplets or confining walls (Sections~\ref{sec:walls} and \ref{sec:droplet_interactions}). 

\begin{marginnote}
\entry{Chemotaxis}{Ability of a self-propelled system to drift or reorient as the result of a local chemical gradient.}
\end{marginnote}
Conversely, a local increase of the bulk concentration in  surfactant monomers will reduce surface tension and generate a locally-diverging Marangoni flow at the droplet's surface, propelling it toward such chemically-rich regions (Figure~\ref{fig:chemotaxis}b). \citet{Jin2017} recently exploited this \emph{chemotaxis} to demonstrate the droplet's maze-solving ability by its sensing of the shortest path toward the surfactant-rich exit (Figure~\ref{fig:chemotaxis}c).

\subsection{Interactions with confining boundaries}
\label{sec:walls}
\begin{figure}
\begin{center}
\includegraphics[width=.95\textwidth]{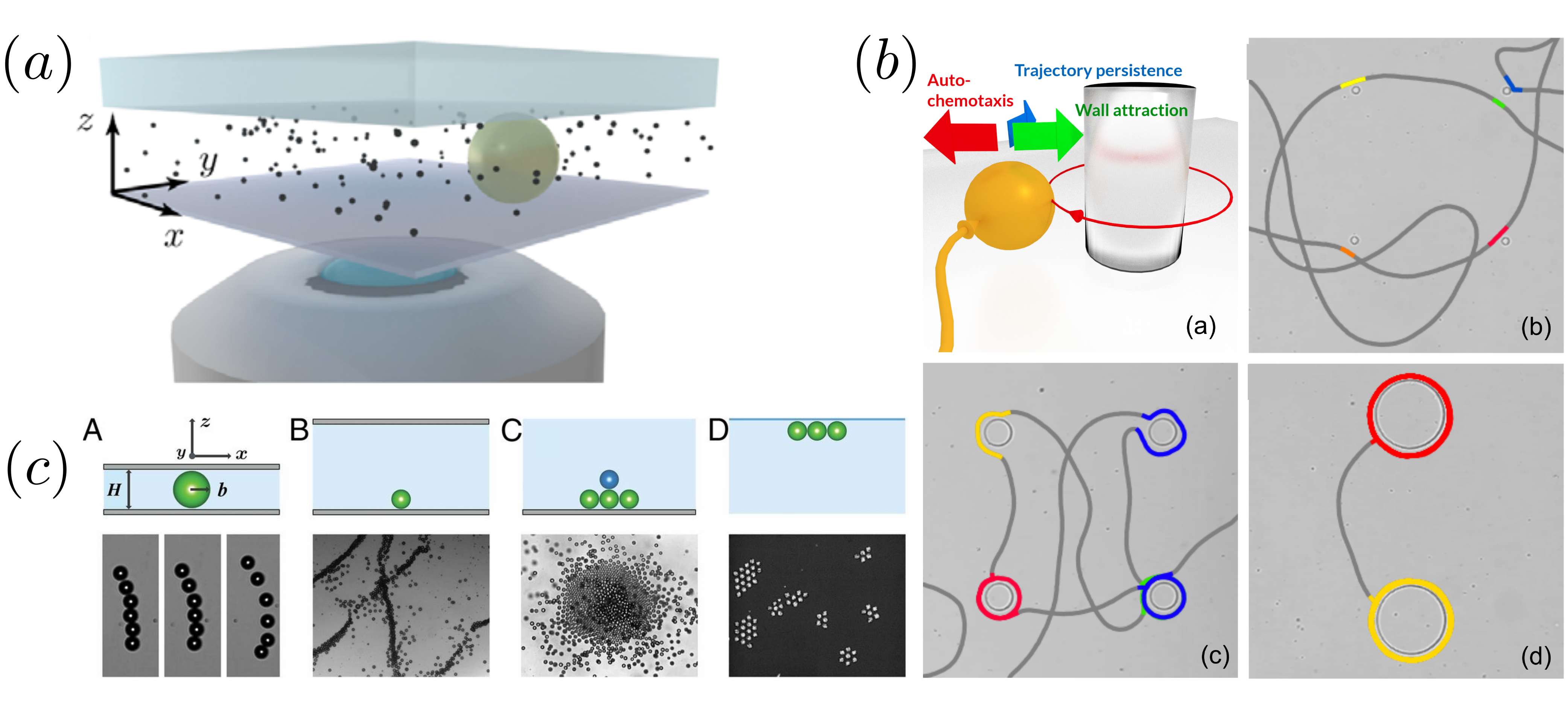}
\caption{Droplet-wall interactions. (a) Swimming droplets are commonly studied in confined environments such as Hele-Shaw cells.  (b) Trapping and guidance of oil droplets around pillars. (c)~The nature and proximity of confining boundaries profoundly modify the collective dynamics and organization of active droplets. Panel a adapted  with permission  from~\citet{Jin2021}; copyright (2021) American Physical Society. Panel b adapted  with permission  from~\citet{Jin2018}; copyright (2018) IOP Publishing\noteTOAR{ ;permission was purchased on CCC}. Panel c adapted  with permission  from~\citet{Thutupalli2018}\noteTOAR{ ;PNAS indicated no copyright note is needed}.}\label{fig:confinement}
\end{center}
\end{figure}

By preventing its diffusion away from the emitting droplet, chemically-inactive boundaries result in a local solute accumulation, which profoundly modifies the flow field generated by the droplet, e.g. when it is maintained near a wall by buoyancy~\citep[][]{Kruger2016b} or  in Hele-Shaw cells~\citep[Figure~\ref{fig:confinement}c,][]{Dey2021}: a wall-normal pumping mode develops and plays a dominant role in the hydrodynamic signature of the droplet~\citep{deBlois2019}.

This accumulation effectively introduces a repulsive droplet-wall interaction, and an approaching droplet will slow down, reverse direction and swim away. Such repulsive interaction~\citep[which is also present for phoretic colloids,][]{Yariv2016} dominate far-field and low-$\Pe$ interactions of active droplets with confining walls; however, the nonlinear hydrochemical coupling allows for non-trivial dynamics when convective effects become dominant in near-field  or larger-$\Pe$ interactions: using the complete model of Sec.~\ref{sec:canonical_model} for axisymmetric collisions, \citet{Lippera2020} demonstrated the dominance of chemical repulsion in the rebound dynamics for low-to-moderate $\Pe$, and the emergence of complex unsteady rebounds for higher $\Pe$, when a fully-nonlinear hydrochemical coupling becomes significant. The similar complete analysis of near-field interactions in oblique collisions remains so far elusive, calling for a more in-depth investigation.

Indeed, swimming droplets are not systematically repelled by interfaces, as attraction by or orbiting around obstacles was also reported~\citep[Figure~\ref{fig:confinement}d,][]{Jin2019}. Such capture is commonly observed for swimming microorganisms, suggesting a possible hydrodynamic origin to this wall attraction~\citep{Hokmabad2022}. Yet, its origin and precise competition with the chemically-induced repulsion remains to be fully understood. Beyond individual dynamics, the presence and nature of a boundary profoundly modifies how active droplets interact with each other~\citep[Figure~\ref{fig:confinement}e,][]{Thutupalli2018}. 

Most active droplets swim in the immediate vicinity of a wall, a feature ignored in most models; a clear understanding of wall-droplet interactions and confined propulsion thus represents one of the most pressing modelling challenges on the topic (Section~\ref{sec:strong_confinement}).

\subsection{Interactions with an external flow}
Chemically-active droplets placed in a flow experience a net drift and reorientation, just like passive colloids~\citep[Faxen's law, see][]{Kim1991}. Combined with self-propulsion, this leads to an active directed motion in non-uniform flows (\emph{rheotaxis}), examples of which abound in active matter, e.g. bacterial rheotaxis~\citep{Marcos2012} or up- or cross-stream migration of Janus phoretic colloids~\citep{Palacci2015,Katuri2018}.

\begin{marginnote}
\entry{Rheotaxis}{Ability of a self-propelled system to   drift or reorient in a local flow gradient (shear).}
\end{marginnote}

In contrast with other systems, the effect of a non-uniform flow on the trajectories of chemically-active droplet remains so far relatively unexplored. \citet{Dey2021} recently reported oscillatory trajectories and upstream migration of isotropic liquid crystal droplets inside a rectangular capillary flow; a simplified far-field model of the flow perturbation introduced by the droplet was shown to reproduce such trajectories, suggesting a purely hydrodynamic origin to this rheotactic behaviour. Yet, this question requires further investigation due to the fundamental inability of far-field models to account for the lubrication effects present for such tightly-confined geometries. Furthermore, \citet{Dwivedi2021b} reported similar dynamics in Couette flows and argued qualitatively instead for a chemical origin to the reorientation of the droplets coupled to the droplet's internal structure. 

More generally, the effect of the flow-induced reorganisation of the chemical distribution around them remains almost entirely unexplored, despite its likely critical role on the droplet propulsion. The asymptotic analysis of the far-field hydrodynamic signature of a model active colloid by a uniform shear flow is a notable exception~\citep{Yariv2017b}, but such analysis remains limited to subcritical $\Pe$, i.e. without any droplet propulsion.

\subsection{Droplet-droplet interactions}\label{sec:droplet_interactions}
\begin{figure}
\begin{center}
\includegraphics[width=.95\textwidth]{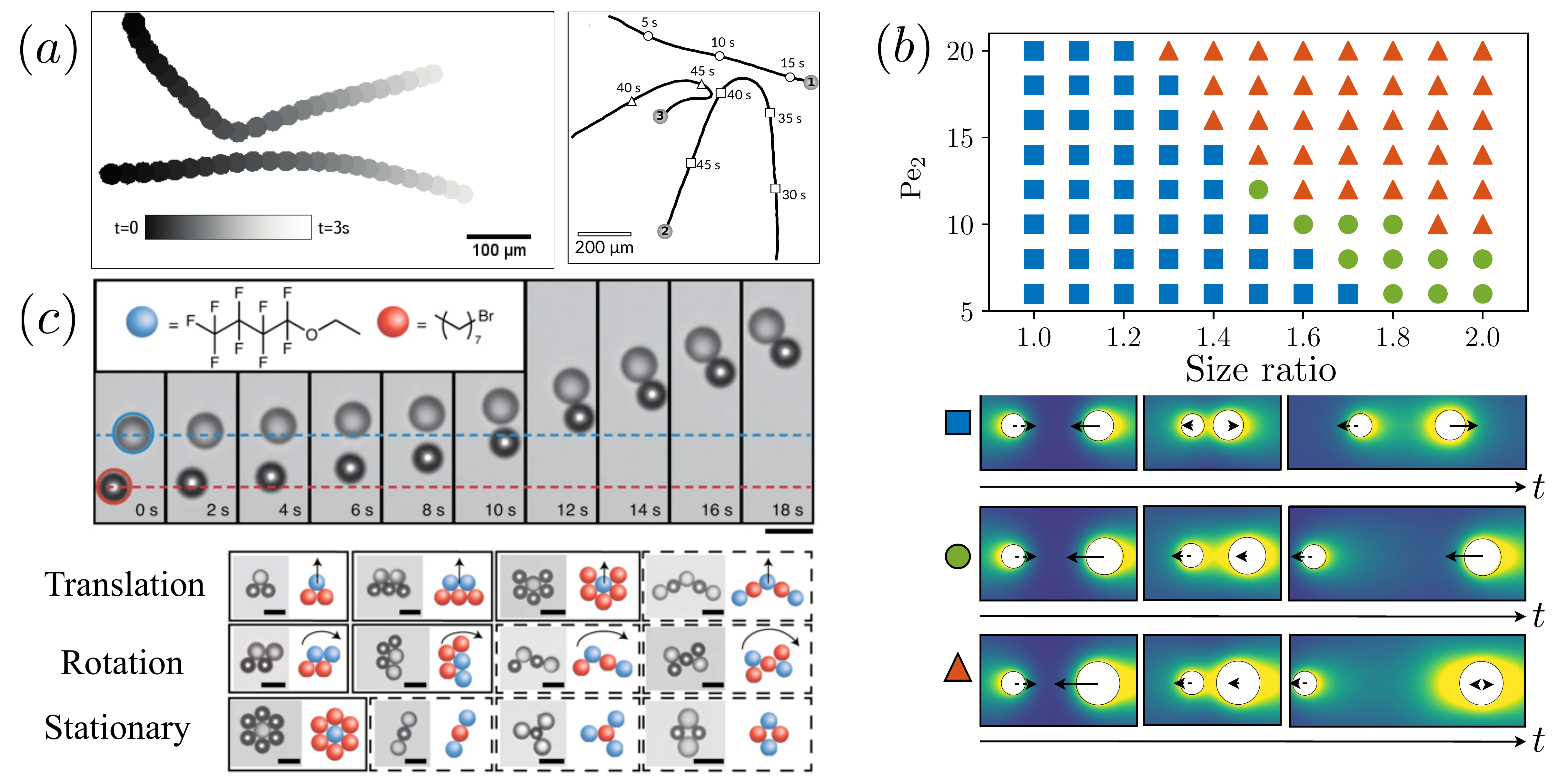}
\caption{Droplet-droplet interactions. (a) Experimental trajectories of water-in-oil (left) or oil-in-water (right) active droplets, which are chemically-repelled by others and their persistent chemical trails. (b) Possible outcome of axisymmetric droplet collisions in numerical simulations of the complete model: depending on their relative size, active droplets undergo asymmetric rebounds (blue square), bound-state chasing regimes (green disk) or mode switching toward a pumping mode for the largest droplet (red triangle). (c) Experimental visualisation of predator-prey interactions (top) or translating, rotating and stationary aggregate formations for droplets of different micellar solubilisation rates. Panel a reproduced  with permission  from~\citet{Moerman2017} (copyright 2017 American Physical Society) and~\citet{Jin2017} \noteTOAR{PNAS indicated that no copyright note is needed}. Panel b adapted  with permission  from~\citet{Lippera2020b}; copyright (2020) American Physical Society. Panel c adapted  with permission  from~\citet{Meredith2020}; copyright (2020) Springer Nature. \noteTOAR{permission was purchased on CCC}}\label{fig:interactions_droplets}
\end{center}
\end{figure}

Self-solubilising droplets are both net sources/sinks of chemical solute and force-free microswimmers and may thus generate $O(1/d^2)$ hydrodynamic and chemical drifts of their neighbours. Such dual hydro-chemical signature influences the trajectories of other droplets, that undergo chemotaxis and hydrodynamic interactions like many swimming cells~\citep{Lauga2020}, resulting in complex individual and collective dynamics (Figure~\ref{fig:interactions_droplets}).

Chemically-active droplet are intrinsically anti-chemotactic, and therefore tend to avoid other droplets or their persistent chemical trails~\citep[Figure~\ref{fig:interactions_droplets}a,][]{Moerman2017,Jin2017}. Yet, the nonlinear  hydrochemical coupling  makes the collision result particularly sensitive to its precise conditions, as illustrated by the outcome diversity and sensitivity of head-on droplet collisions to small droplet size contrasts~\citep[see Figure~\ref{fig:interactions_droplets}b,][]{Lippera2020b}.
A complete hydro-chemical modeling of more generic oblique collisions remains so far elusive, and \citet{Lippera2021} proposed instead a reduced-order model based on the dominant chemical interactions and the finite-time reorientation of the droplet's wake, predicting and characterising the strong sensitivity of the collision outcome (alignment or scattering) to the droplet's initial arrangement. These results were then recently confirmed by \cite{Hokmabad2020b}, further providing unique visualisations of the droplets' chemical trails. Long-ranged repulsive chemical interactions can however not explain the droplets' long-term alignment, whose hydrodynamic origin remains to be confirmed~\citep{Hokmabad2022}. In the long-term, the persistence of the droplets' repelling chemical trails profoundly constrains the droplet trajectories and their long-time dynamics, introducing an effective ``cageing'' effect~\citep{Hokmabad2020b}.

Even more diverse interactions arise from variability in chemical composition: droplets of different relative solubility  or solubilisation rates may effectively use each other's wake as fuel leading to peculiar and tunable predator-prey dynamics, as well as assembly into larger droplet clusters with distinct dynamical properties~\citep[Figure~\ref{fig:interactions_droplets}c,][]{Meredith2020}.

\subsection{Forced motion of active droplets}
\begin{figure}
\begin{center}
\includegraphics[width=.95\textwidth]{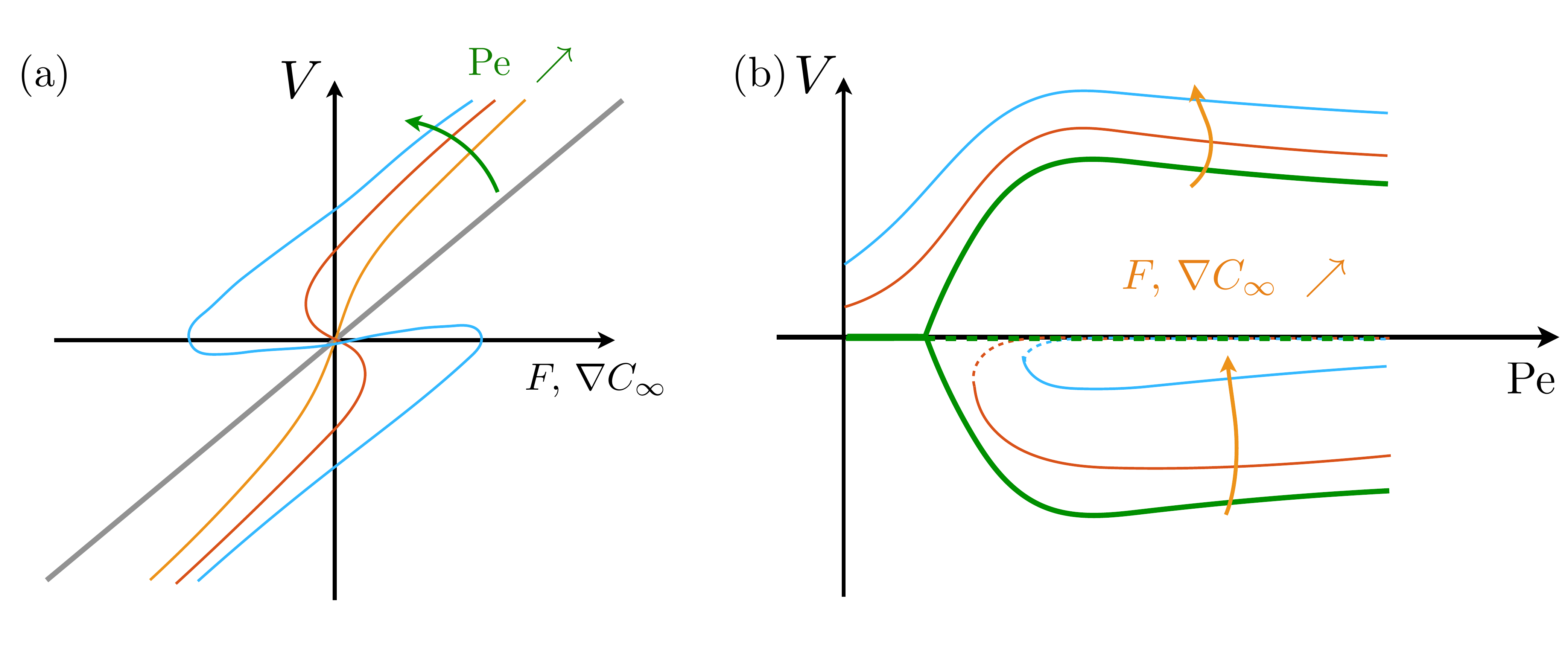}
\caption{Spontaneous motion of a forced droplet obtained numerically from the reference nonlinear hydro-chemical model (Sec.~\ref{sec:canonical_model}). (a) Forcing-velocity diagram. (b) Modified bifurcation diagram under the effect of a directional external forcing (external force or background concentration gradient), which intrinsic isotropy of the problem. Stable and unstable branches are respectively denoted as solid and dashed lines. Motion always occurs in the direction induced by the forcing (upper branch), but propulsion against the forcing is also possible for sufficiently large $\mbox{Pe}$.\noteTOAR{this figure was created for this review}}\label{fig:bifurcation}
\end{center}
\end{figure}

The symmetry-breaking instability and emergence of self-propulsion can be generalised to active droplets experiencing a net non-hydrodynamic force $F$ (e.g. gravity). This is in fact particularly relevant experimentally as droplets are typically not neutrally-buoyant. A chemically-passive droplet experiences a net linear drift in response to an outer force (Stokes law).
For a chemically-active droplet (i.e. $\Pe>0$), the forcing-velocity relation becomes nonlinear (Figure~\ref{fig:bifurcation}a); for imposed $F\neq 0$, the Marangoni forcing resulting from the accumulation of solute in the wake of the drifting droplet enhances self-propulsion, i.e. the effective hydrodynamic drag is reduced and velocity-dependent~\citep{Yariv2017b}. As $\mbox{Pe}$ is increased further, we note the emergence of (i)~\emph{self-propelled force-free droplets} ($F=0$, $V\neq 0$) for $\mbox{Pe}>\Pe_c^{(1)}$ corresponding to spontaneous motion, and (ii)~\emph{non-swimming pumping droplets} ($V=0$, $F\neq 0$) for $\Pe>\Pe_c^*>\Pe_c^{(1)}$ that stir the fluid to maintain the required concentration contrast and Marangoni forcing balancing the net outer forcing~\citep[see also][]{Rednikov1994b}.

This suggests an imperfect bifurcation~\citep{Saha2021}, as illustrated in Figure~\ref{fig:bifurcation}b for axisymmetric propulsion. For $F=0$, self-propulsion occurs spontaneously in any direction for $\Pe>\Pe_c^{(1)}$; an external (directional) force breaks the symmetry of the two branches. Stable propulsion \emph{along} the outer force is always observed; for fixed $\Pe$ and sufficiently small $F$, two other solutions are found, including a stable motion \emph{against} the imposed forcing, a scenario consistent with recent reports of droplet motion against gravity~\citep{Moerman_thesis}.

A similar framework applies to the force-free motion of active  droplets in externally-imposed chemical gradients, as that generated by neighbouring droplets~\citep{Rednikov1994d,Lippera2020}; the modified hydrodynamic footprint associated with the different forcing, and its impact on chemical transport and induced Marangoni stresses must however be accounted for carefully.

\section{Perspectives: some open problems}
\label{sec:perspectives}
In addition to the challenges already presented (e.g. large $\Pe$ dynamics), we conclude this review by a brief overview of some open questions in the characterisation and modeling of swimming droplets, and of key ongoing contributions to resolving them.

\subsection{Strongly confined droplets}

\label{sec:strong_confinement}

In most experiments, active droplets swim close to and along rigid walls~\citep{deBlois2019,Cheon2021}, within Hele-Shaw cells~\citep{Hokmabad2021} or even small capillaries~\citep{deBlois2021}, offering a clear demonstration that large viscous stresses classically associated with lubrication layers do not hinder self-propulsion but  alter their collective dynamics~\citep{Kruger2016b,Thutupalli2018}.
Yet, ignoring such intrinsic experimental features, existing models focus almost exclusively on propulsion in \emph{unbounded} environments with slow (algebraic) far-field decays and long-ranged interactions. Accurate modeling of confined propulsion is thus one of the main challenges of the field. 

Partially lifting the veil, \citet{Desai2021} recently demonstrated using the reference model, that a nearby wall actually promotes the symmetry-breaking instability and onset of propulsion. Stronger confinement further results in enhanced swimming speeds~\citep{Picella2022}, consistently with recent experimental observations inside tight capillaries~\citep{deBlois2021}. Both of these theoretical studies point to two critical features of confined motion: (i) a fundamental modification of the solute's transport  around and away from the droplet to cope with the wall-restricted solute diffusion, and (ii) the localization of driving interfacial stresses to the most confined regions where hydrodynamic resistance (i.e. grip on the substrate) is largest. 

\subsection{Chemical dynamics, restricted diffusion and mathematical well-posedness}\label{sec:wellposed}
The reference model has the clear advantage to retain the critical physical ingredients for droplet propulsion while remaining minimal and conceptually simple. But its underlying mathematical characteristics should not be overlooked. In fact, understanding whether potential peculiar behaviours result from modeling shortcomings  or are an intrinsic  feature of the physico-chemical problem is a fundamental challenge, and the well-posedness of the chemical transport problem provides an enlightening illustration. 

Absent any bulk chemical relaxation, the existence of a steady state solution (propelling or not) requires transporting the released solute away from the droplet's surface. As for the classical problem of a sedimenting heat/mass-releasing sphere~\citep{Acrivos1962}, a fundamental dichotomy arises between diffusion-dominated transport at $O(R)$-distances, and a convection-driven region at $r=O(D/V)$. When $VR/D\ll 1$, these two scales are clearly separated, introducing a far-field boundary layer whose existence in fact conditions the bifurcation and onset of self-propulsion~\citep{Velarde1996,Farutin2021a}.

Finite-$\Pe$ self-propulsion of homogeneous active droplets (or phoretic particles) develops from an isotropic state with no fluid or droplet motion and where the solute can only be transported away by diffusion. Such balance  is possible in $3D$, albeit precariously,  but breaks down as soon as chemical transport is geometrically-restricted (e.g. in 2D or in a channel) and a regularizing mechanism is required, such as time-dependence~\citep{Sondak2016} or convection~\citep{Yariv2017}. Such technical yet fundamental difficulties are not present when the solute concentration spontaneously relaxes to equilibrium in the bulk fluid~\citep[e.g.][]{Yoshinaga2012}, which introduces an exponential chemical screening.

\subsection{Coupling with the droplet internal dynamics and control opportunities}
The fluid nature of the droplet is enough to understand the spontaneous onset of its motion or some of its complex dynamics, but its internal structure may introduce critical changes and controls that represent  important challenges and opportunities for the field.  The presence of a topological defect~\citep{Kruger2016a} or an internal core~\citep[][]{Hokmabad2019} in liquid crystal droplets or shells can promote curly, helicoidal or meandering trajectories, as a result of the coupling of the droplet motion with an internal \emph{polarity}~\citep{Morozov2019c}. Helical propulsion can also be directly achieved with non-nematic droplet~\citep{Hokmabad2022} or using the internal chirality of a cholesteric liquid crystal droplet~\citep{Yamamoto2017}. Encapsulating solid particles in the droplet enhances the self-propulsion velocity by providing yet another internal polarisation of the droplet structure~\citep{Cheon2021}. Multi-component droplets also gain, at least transiently, an internal polarity due to the differential dissolution and phase separation of their individual components~\citep{Li2020,Wang2021}.

\begin{marginnote}
\entry{Topological defect}{Point or line singularity in the local directional ordering of a liquid crystal.}
\end{marginnote}

Beyond the need and interest of a fundamental understanding, this influence of the droplet properties on its motion provides control opportunities. Incorporating chiral molecular motors into a nematic liquid crystal, \citet{Lancia2019} prepared cholesteric droplets whose helical trajectories and their handedness can be controlled by an external light. In a similar spirit, light-controlled switches of the conformation of the surfactant covering the droplet surface can be used to change the droplet solubilisation rate and produce run-and-halt motions~\citep{Ryabchun2021}. The sensitive conditions for the internal kinetics of chemically-reacting droplets can also be used to control the propulsion mode, such as the pH of the outer fluid~\citep{Ban2013}, or the internal state of the oscillatory BZ reaction~\citep{Kitahata2011,Thutupalli2013,Suematsu2021}.

\begin{summary}[SUMMARY POINTS]
\begin{enumerate}
\item Chemically-active droplets emit, consume or alter surfactant monomers or micelles; they exploit the nonlinear convective transport by the surface-driven flows resulting from these solutes' gradients to self-propel.
\item Due to this nonlinear coupling and their isotropy, their individual motion is fundamentally different from other (asymmetric) chemically-active colloids.
\item Their dual hydro-chemical footprint leads to complex interactions with other droplets or confining boundaries.
\end{enumerate}
\end{summary}

\begin{issues}[FUTURE ISSUES]
\begin{enumerate}
\item What are the origin and characteristics of the individual complex trajectories and chaotic-like dynamics of isolated droplets observed in experiments?
\item How can we model the dynamics of many active droplets or within complex geometries, while retaining the intrinsic nonlinearity of the hydrochemical coupling? Are direct numerical simulations the only resort? If so, what are the most adapted methods to accurately treat the nonlinear coupling and moving droplet boundaries?
\item What is the influence of their geometric environment on the motion of active droplets? How relevant are bulk models to understand observations?
\item Existing models put much focus on hydrodynamic and transport modelling. Are the details of the physical chemistry involved important? What is the role and features of the surfactant dynamics at the densely covered interface?
\item Can the motion of active droplets be controlled? If so  how and for which purposes can they realistically be used for engineering applications? 

\end{enumerate}
\end{issues}
\section*{DISCLOSURE STATEMENT}
The author is not aware of any affiliations, memberships, funding, or financial holdings that
might be perceived as affecting the objectivity of this review. 

\section*{ACKNOWLEDGMENTS}
I am grateful to Eric Lauga and Cecile Cottin Bizonne for their careful feedback on the manuscript, to Olivier Dauchot for providing some of the experimental photos included here, and all of them as well as Denis Bartolo, Matvey Morozov, Ory Schnitzer, Udi Yariv and Jasna Brujic for many insigtful discussions on the dynamics of active droplets. 
This work has received funding from the European Research
Council (ERC) under the European Union’s Horizon 2020
research and innovation programme (Grant agreement
714027)
%




\end{document}